# Dynamics of the variation of shape of the hysteresis of stress-strain cyclic compression curves registered at the investigations of the mechanical behavior of composite hydrogels

I.V. Gofman*, A.L. Buyanov

Institute of Macromolecular Compounds, RAS, St.-Petersburg, Russia

An unusual type of mechanical behavior was registered while studying the swollen hydrogel compositions "cellulose-polyacrylamide" in the conditions of multiple cyclic compression tests with the broad variation of the deformation speed. While increasing the deformation speed the clearly seen inversion of the positions of compression and decompression parts of the cyclic stress-strain curves was detected. While carrying out the cyclic compression tests with relatively low deformation speed (about 100-200 % of the initial sample's height per minute) the well defined hysteresis of the stress-strain curve can de seen and in these conditions the decompression part of the curve is situated inferior the part corresponding to compression. But while increasing the speed of the deformation the tendency to the progressive approach of the compression and decompression curves to each other is clearly seen. This effect results in the full disappearance of the hysteresis at some value of the deformation speed: the decompression curve coincides with the compression curve. Along with the further increase of the deformation speed the hysteresis appears again but the curve corresponding to compression process is situated inferior the curve corresponding to decompression: the "inversion" of the hysteresis was detected. The precise character of this process depends upon the stiffness of the hydrogel under study. Up to date the convincing explanation o this effect can not be put forward. The authors can only present some hypotheses to explain this phenomenon.

* Corresponding author: Institute of Macromolecular Compounds, Russian Academy of Sciences, Bolshoy st., 31, St.-Petersburg, 199004, Russia. Tel.: +7(812)328-8511.
E-mail addresses: gofman@imc.macro.ru (I.V. Gofman), buyanov799@gmail.com (A.L. Buyanov).

During the last decade, research on the synthesis and characterization of polymer hydrogels has became one of the extensively developing branches of polymer science. These polymer systems were shown to be the promising materials for biomedical applications, for example as the artificial cartilage to substitute the injured natural one [1-3]. Indeed, different types of polymer hydrogels are very close to human tissues by the structure and some physical properties. Moreover these materials are not toxic at all. Under these reasons they possess the remarkable biological compatibility. However, from the viewpoint of the mechanical properties, first of all - the mechanical stiffness, only some special types of hydrogels are close to articular cartilages [4, 5]. Up to date the problem of improving the mechanical properties of the hydrogel materials to meet the requirements to the properties of the artificial substituent of natural articular cartilage has not been completely solved [6, 7], and this situation hampers the real use of hydrogels in this branch of medicine. The systematic data concerning the peculiarities of the visco-elastic behavior of such hydrogels in comparison with that of natural cartilages can hardly been found in the scientific literature.

Our team is carrying out the extensive investigations of these materials for several years. During this work the new types of composite hydrogels were elaborated: the materials with the structure of interpenetrating networks of cellulose and polyacrylamide (C–PAAm) [4, 5]. These materials with the high degree of biological compatibility are maximally close to the natural cartilage by their mechanical properties [8]. Under these reasons the hydrogels of this type can be successfully used as the artificial analogs, substituents of human cartilage of different localization, first of all the joints cartilage.

The properties of these materials can be substantially varied depending upon the hydrogel quantitative composition and the peculiarities of the synthesis. The protocol of the synthesis of the materials under consideration was described in our previous works [4, 8].

In the course of these investigations the close attention was paid to the mechanical properties of the materials under study because the cartilage in human joint is subjected to the permanent action of the substantial mechanical stresses. These properties were studied in different test conditions and in different deformation modes.

To characterize the peculiarities of the mechanical behavior of these materials in the loading conditions close to the working conditions of joint cartilage we are widely using both single-shot compression tests (Fig. 1) and different types of cyclic compression tests [5, 8]. The most widely used mode of these tests is the multiple cyclic compressions of the cylindrical hydrogel samples with constant amplitude of the deformation as high as 30-50 %. These values of amplitude correspond to the range of maximal values of the compression of the human knee cartilage in different modes of movement [9-11]. The tests of the hydrogels under study in these conditions are most suitable to clarify the extent of the stability of the behavior of these materials under the action of the long term mechanical loads.

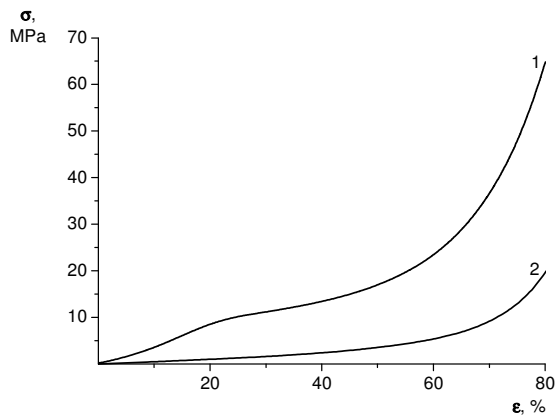

Fig. 1. Stress-strain curves of the stiff and soft hydrogel samples obtained in the conditions of single-shot compression tests.

At these tests the cylindrically shaped samples of swollen hydrogel (5-7 mm height and ~10 mm diameter, Fig. 2) are subjected to the multiple unconfined compression acts with a constant deformation amplitude (30-50 %) followed by the decompression with the same speed up to the initial height of the sample. Depending upon the goals of each test its duration can vary in a broad range of cycles' number – from several cycles up to several thousands of cycles.

The protocol of these tests was described in details in [5, 8].

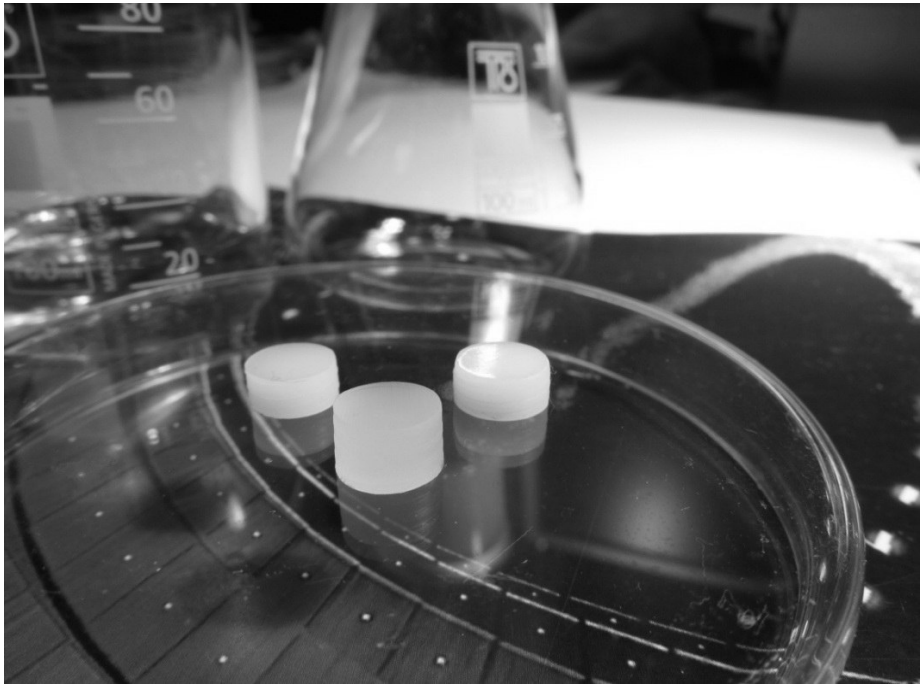

Fig. 2. The hydrogel samples used in our tests.

During all our previous works these cyclic tests were carried out at the same deformation speed of 10 mm/min which corresponds to ~200 % of the initial sample's height per minute [5]. The typical cyclic stress-strain curves obtained under these test conditions at the hydrogels of different stiffness are presented at Fig. 3a, and 4a.

But at the next step of our investigations we have tried to study the cyclic compression behavior of these materials in the conditions of the tests with different deformation speeds. The hydrogel materials of different stiffness denoted as "stiff gel" and "soft gel" (both based on modified plant cellulose [4]) were subjected to these tests. The data characterizing the extent of swelling and the stiffness of these materials in the conditions of one-shot unconfined compression (Fig. 1) are presented in Table 1. The detailed protocol of these single compression tests was described in [5].

The following characteristics are presented:

- the water content in the material in the equilibriously swollen state

W;

- the mean values of compression modulus at two deformation ranges - 10-15 % and 25-30 %, $E|_{10\text{-}15\%}$ and $E|_{25\text{-}30\%}$, respectively (the modulus values were calculated as ratios of stress increments $\Delta\sigma$ to strain increments $\Delta\varepsilon$ in the appropriate ranges of deformation);

- the stress corresponding to the specimen's compression of 80 % - $\sigma_{80\%}$.

Table 1. Mechanical characteristic of hydrogel samples under study

| Hydrogel type | W, % | $E|_{10\text{-}15\%}$, MPa | $E|_{25\text{-}30\%}$, MPa | $\sigma_{80\%}$, MPa |
|---|---|---|---|---|
| Soft gel | 75 | 1.5 | 3.2 | 19.8 |
| Stiff gel | 52 | 21.4 | 105.6 | 66.3 |

In the cyclic tests we first of all were waiting for the variations in the stiffness of the hydrogels provoked by the variation of the deformation speed. This effect would reflect the visco-elastic nature of the mechanical behavior of these materials. As a result the increase in the stresses could be registered at the cyclic compression while increasing the deformation speed.

To prove this assumption we have carried out the cyclic tests (50 cycles in each test) of both stiff and soft gel samples with the speed values varied from 1 to $1\times10^3$ mm/min (from ~20 to ~$2\times10^4$ % of the initial sample's height per minute). The test at each new speed was carried out using a new, previously unused sample. During the tests the hydrogel samples were put in water medium. The examples of cyclic stress-strain curves characterizing the behavior of both stiff and soft gels at different deformation speeds are presented at Fig. 3 and Fig. 4, respectively. It should be noted that the stabilization of the shape of these cyclic stress-strain curves of both hydrogels under study takes place already after the first compression cycle. In Fig. 3-4 the 5-th cycles are presented.

The cyclic stress-strain curves obtained at the low deformation speeds are characterized by the well defined hysteresis (Fig. 3a-4a). Namely the stresses at the compression parts of the cycles are higher than those at the decompression parts at the same deformation values. For stiff hydrogel this hysteresis can be seen already if the deformation amplitudes are as high as 25-30 % (Fig. 3), and while cycling the soft gel it can be clearly registered at more substantial deformation amplitudes – as high as 45-50 % (Fig. 4).

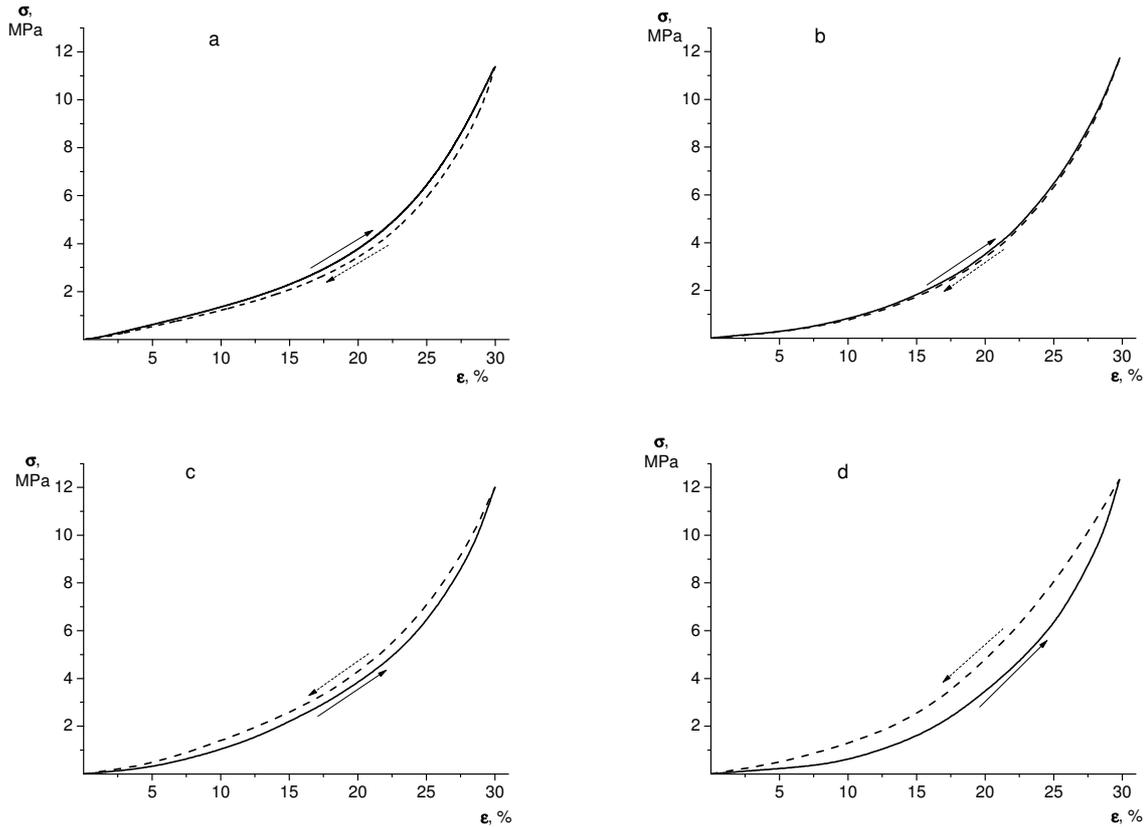

Fig. 3. Cyclic stress-strain curves of the compression of stiff gel (see Table 1) with different deformation speeds: a – 10 mm/min, b – 100 mm/min, c – 300 mm/min, d – 500 mm/min.

The first conclusion that can be drawn from our tests is that only little variations of the stiffness of both types of hydrogels studied were detected under the variations of the deformation speed in a broad (~3 orders of magnitude) range of its values. Indeed, for the stiff gel the maximal compression stress value at the cycle increases from 11.27 to 12.33 MPa while the deformation speeds was increased from 1 to 500 mm/min (amplitude of the compression was 30 %, Fig. 3), and for the soft gel this value varies from 2.88 to 3.58 MPa while the deformation speeds was increased from 1 to 1000 mm/min (amplitude of the compression was 50 %, Fig. 4).

This character of the behavior of the materials studied is close to that of the articular cartilage presented in [12].

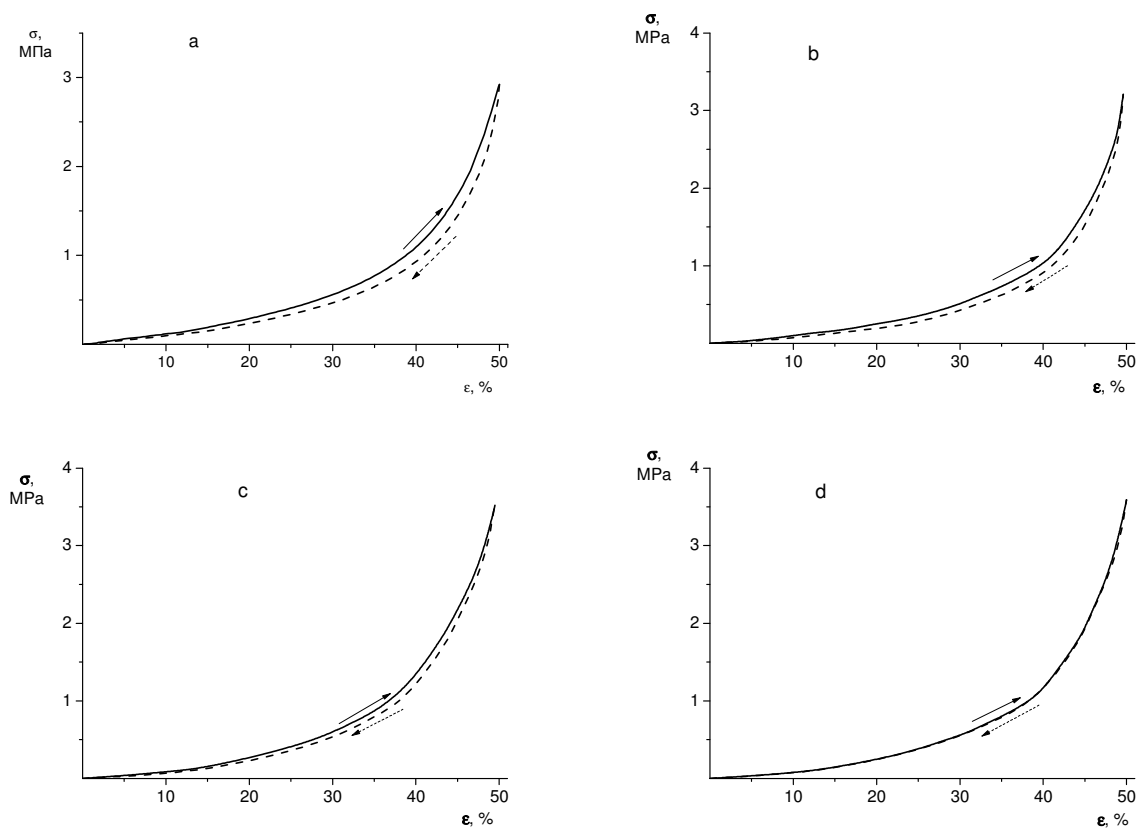

Fig. 4. Cyclic stress-strain curves of the compression of soft gel (see Table 1) with different deformation speeds: a – 10 mm/min, b – 200 mm/min, c – 500 mm/min, d – 1000 mm/min.

But one unusual effect was registered during these tests.

While increasing the deformation speed the clearly seen inversion of the positions of compression and decompression stress-strain curves of the cycle was detected for the stiff gel (Fig. 3). During the increase of the deformation speed up to ~100 mm/min the hysteresis becomes more and more weak, the compression and decompression parts of the cyclic stress-strain curve tends to move to each other. This process results in practically full disappearance of the hysteresis of the material's cyclic stress-strain behavior in the deformation speeds range 100-200 mm/min ((2-4)×10$^3$ % of the initial sample's height per minute). At higher deformation speeds the real inversion of the positions of two parts of the cyclic stress-strain curve can be seen (Fig. 3c-3d): stresses at decompression are higher than those at compression at the same deformation values, and the increase in the deformation speed leads to the broadening of this inverse hysteresis, to the increase of its area.

While testing the soft gel (Fig. 4) the inversion of the positions of compression and decompression parts of stress-strain curves under the variations of the deformation speed was not clearly evidenced up to the maximal speed realized in our tests (1000 mm/min). Only the tendency to a gradual decrease of the area of the hysteresis can be seen: the compression and decompression parts of the cyclic stress-strain curve tends to move to each other. This process leads to the disappearance of the hysteresis that was registered at the maximal deformation speeds - 800-1000 mm/min (Fig. 4d). Apparently the inversion of shape of the hysteresis can take place for these soft gels at the deformation speed yet higher than the maximal speed that we can realize in our equipment.

To make sure of the real physical nature of the effect described, to bar some artifacts provoked by the incorrect work of experimental equipment, we have repeated our cyclic tests using the mechanical test systems of three different types: UTS 10 (UTStestsysteme, Germany), Instron 3365 (Instron), and AG-Xplus (Shimadzu Corp., Japan). In all cases we have obtained the same results.

Studying the peculiarities of this phenomenon we tried to answer the question: what part of the cyclic curve, compression curve or decompression one changes chiefly its curvature to insure the reported inversion of the hysteresis? To clarify the situation we have traced the dynamics of curvature variations of both parts of the cyclic compression-decompression curves of the hydrogels studied against the variation of the deformation speed. This process can be characterized by the variations of the compression and decompression modulus values $E|_{10-15\%}$, caused by the increase in the deformation speed. The results obtained showed that both parts of the cyclic curve take part in this process: the curvature of the compression curve increases while the curvature of the decompression curve decreases successively along with the increase of compression-decompression speed (Table 2, the results for the stiff gel are presented as an example).

Table 2. Variations of the mean curvatures of the compression and decompression parts of the cyclic stress-strain curves of the stiff gel vs. the deformation speed

| Test speed, mm/min | $E\|_{10-15\%}$,* MPa | | |
|---|---|---|---|
| | compression | decompression | mean value |
| 1 | 21.5 | 19.7 | 20.6 |
| 10 | 21.2 | 19.9 | 20.6 |
| 100 | 21.1 | 20.8 | 20.9 |
| 200 | 20.9 | 21.4 | 21.2 |
| 300 | 20.6 | 22.4 | 21.5 |
| 500 | 18.9 | 24.8 | 21.9 |

* See the explanations to Table 1

Up to date we can not put forward any reasonable explanation of this behavior of the materials under study. It may reflect apparently the relations of the deformation speed and the characteristic times of the mechanical relaxation processes in different components of the materials under study, and in the whole material's structure. The interactions of the polymers' macrochains of hydrogel with water containing in its volume should be taken into account too while trying to understand the origin of the unusual behavior described above. The computer simulation of the situation is very difficult because of a wide set of relaxation processes which can take place in these swollen materials with the structure of interpenetrating networks.

The interpretation of the experimental results presented above will be the task of our further work in this direction.

**Author Contributions**

Both authors contributed equally.

**Acknowledgements**

The work was supported by the grant from the Program of the Presidium of the Russian Academy of Sciences «Fundamental research for the development of biomedical technologies» (FIMT-2014-066).